\documentclass[prl,amsmath,amssymb,superscriptaddress,twocolumn]{revtex4}
\usepackage{graphicx}
\usepackage{bbm,color,ulem,mathbbol}
\newcommand{\ba}{{\bf a}}

\newcommand{\br}{{\bf r}}
\newcommand{\bu}{{\bf u}}
\newcommand{\bx}{{\bf x}}
\newcommand{\by}{{\bf y}}
\newcommand{\bk}{{\bf k}}

\newcommand{\bz}{{\bf z}}
\newcommand{\bq}{{\bf q}}
\newcommand{\bA}{{\bf A}}

\newcommand{\bK}{{\bf K}}

\newcommand{\bR}{{\bf R}}
\newcommand{\bdelta}{\delta}

\DeclareMathAlphabet{\mathpzc}{OT1}{pzc}{m}{it} 
\begin{document}
\title{Anomalous scaling and gapless fermions
of $d$-wave superconductors in magnetic field}
\author{Oskar Vafek}
\affiliation{National High Magnetic Field Laboratory and Department
of Physics,\\ Florida State University, Tallahassee, Florida 32306,
USA}
\date{\today}
\begin{abstract}
The problem of $d_{x^2-y^2}$-wave quasiparticles in a weakly
disordered Abrikosov vortex lattice is studied. Starting with a
periodic lattice, the topological structure of the magnetic crystal
momenta of gapless fermions is found for the particle-hole symmetric
case. If in addition the site centered inversion symmetry is
present, both the location and the number of the gapless fermions
can be determined using an index theorem. In the case of spatially
aperiodic vortex array, Simon and Lee scaling is found to be
violated due to a quantum anomaly. The electronic density of states
is found to scale with the root-mean-square vortex displacement as
$\sqrt{H}f(\bu_{rms}^2/\xi^2)$, while thermal conductivity is
$H$-independent, but {\it different} from the $H=0$ case.
\end{abstract}
\maketitle
\label{section:introduction}

Theoretically, it is well established that the motion of the low
energy fermionic quasiparticles in the Abrikosov vortex state of
$d-$wave superconductors is determined by the combination of two
effects: the so-called Doppler shift\cite{Volovik1993} and the $Z_2$
Berry phase\cite{FranzTesanovic2000}. The former arises from the
coupling of the quasiparticle charge current to the superfluid
velocity and has a classical analog, while the latter is purely
quantum and corresponds to an extra $\pi$ Aharonov-Bohm phase
accumulated by the quasiparticle (qp) wavefunction upon encircling a
vortex. Semiclassical analysis of the Doppler shift coupling lead
Volovik \cite{Volovik1993} to the conclusion that the original nodal
spectrum of a $d$-wave superconductor should be filled upon the
application of the external magnetic field, $H$, resulting in the
finite density of states at zero energy increasing as $\sqrt{H}$.
Experimental data on, for example, specific heat and thermal
conductivity have been largely interpreted in this way.

However, the relatively recent experiments on the ultraclean
YBCO\cite{Hill2004} have failed to detect the $\sqrt{H}$ dependence
of the qp contribution to the low temperature thermal conductivity
$\kappa_{xx}$. Instead, in the sub Kelvin temperature regime
$\kappa_{xx}/T$ approaches a value which is essentially $H$
independent\cite{Hill2004}. This is unlike the less pure YBCO
samples where the said $\sqrt{H}$ dependence was observed. It is
worth stressing that the qp mean free path, which has been estimated
in Ref.\cite{Ong2001}, exceeds the magnetic length for the $H$
fields above 1Tesla by several orders of magnitude. This suggests
that, in the ultraclean samples, the dominant qp scattering in the
Abrikosov state comes from the {\it positional vortex disorder},
rather than from the impurity scattering. This in turn implies, that
the quantum effects play an important role.

In this work I suggest a theoretical approach to calculate the
scattering rate due to the weak positional vortex disorder as well
as the resulting thermal conductivity. The key idea is to {\it
exactly} solve the periodic lattice problem for a vortex lattice
with a site centered inversion symmetry and a particle-hole
symmetry, along the lines of Ref.\cite{VafekMelikyan2006}, and then
use the symmetry of the resulting nodal qp wavefunctions to find the
effects of various perturbations. In this way one can analyze quite
rigorously the effects of weak positional vortex disorder. The new
finding is that the coupling of the small displacement $\bu_{\bR}$
of a vortex at $\bR$ to the {\it new} vortex lattice nodal
quasiparticles\cite{VafekMelikyan2006} can be thought of as a
coupling to a small vortex-antivortex dipole. Due to the anomalous
$r^{-\frac{1}{2}}$ increase of the exact qp wavefunctions near a
vortex\cite{VafekMelikyan1_2001,Melikyan2006}, this coupling is
anomalously large and enters the effective Hamiltonian via the
dimensionless ratio $\bu_{\bR}/\xi$ where $\xi$ is a short distance
cutoff of the order of the core radius, rather than via
$\bu_{\bR}/\ell$ as suggested by the scaling
argument\cite{SimonLee1997,Marinelli2000}. Weakly disordered vortex
lattice is then viewed as a collection of small vortex-antivortex
dipoles with random magnitude and random orientation whose effect on
the new nodal qp can be analyzed within the self-consistent Born
approximation. The resulting vortex lattice disorder-induced density
of states is found to be finite at zero energy, with an anomalous
dependence on the root-mean-square fluctuation of the vortex
displacement $\bu_{rms}$, scaling as $\sqrt{H}f(\bu^2_{rms}/\xi^2)$,
which is {\it parametrically larger} than in the absence of the
anomaly. For small $\bu_{rms}$, the low temperature thermal
conductivity is found to be very weakly $H$-dependent, in
qualitative agreement with the experimental findings of
Ref.\cite{Hill2004}.

The starting point is the lattice Bogoliubov-de~Gennes (BdG)
eigenequation\cite{deGennes1966}
\begin{eqnarray}
\mathcal{\hat{H}}_0\psi_{\br}=E_n\psi_{\br}\textcolor{blue}.
\end{eqnarray}
where the Hamiltonian $\mathcal{\hat{H}}_0$ acts on the two
component Nambu spinor $\psi_{\br}=[\mathpzc{u}_{\br},
\mathpzc{v}_{\br}]^T$ and has the following explicit form
\begin{equation}\label{bdg0}
\mathcal{\hat{H}}_0= \left(\begin{array}{cc}
\mathcal{\hat{E}}_{\br}-\mu & \hat{\Delta}_{\br}  \\
\hat{\Delta}^{\ast}_{\br} & -\mathcal{\hat{E}^{\ast}}_{\br}+\mu
\end{array}\right).
\end{equation}
Both  $\mathcal{\hat{E}}_{\br}$ and $\hat{\Delta}_{\br}$ are defined
through their action on a lattice function $f_{\br}$ as
\begin{eqnarray}
\mathcal{\hat{E}}_{\br}f_{\br}&=&-t\sum_{\bdelta=\pm\hat{\bx},\pm\hat{\by}}
e^{-i\bA_{\br\br+\delta}} f_{\br+\delta},\\
\hat{\Delta}_{\br}f_{\br}&=&\Delta_{0}\sum_{\bdelta=\pm\hat{\bx},\pm\hat{\by}}
e^{i\theta_{\br \br+\bdelta}}\eta_{\bdelta}f_{\br+\delta}.
\end{eqnarray}
In the symmetric gauge, the magnetic flux $\Phi$ through an
elementary plaquette enters the Peierls factor via
$\bA_{\br\br+\hat{x}}=-\pi y \Phi/\phi_0$, $\bA_{\br\br+\hat{y}}=\pi
x \Phi/\phi_0$; the electronic flux quantum is $\phi_0=hc/e$. The
$d$-wave symmetry is encoded in $\eta_{\bdelta}=+(-)$ if $\bdelta
\parallel \hat{\bx}(\hat{\by})$.
Orthorhombic distortions can be modeled by making horizontal bonds
slightly different from the vertical ones, none of which will change
the key results.  The main topological feature of $\theta_{\br\br'}$
is its $2\pi$ winding around the magnetic field induced vortices.
The initial Ansatz\cite{VafekMelikyan2006} for the pair phases is
\begin{eqnarray}\label{phase}
e^{i\theta_{\br\br'}}\equiv(e^{i\phi_{\br}}+e^{i\phi_{\br'}})/|e^{i\phi_{\br}}+e^{i\phi_{\br'}}|,
\end{eqnarray}
where $\nabla\times\nabla\phi(\br)=2\pi
\hat{\bz}\sum_{i}\delta(\br-\br_i)$ and
$\nabla\cdot\nabla\phi(\br)=0$ where $\br_i$ denotes the vortex
positions.

Connecting pairs of vortices by branch cuts\cite{VafekMelikyan2006},
we can define the singular gauge
transformation\cite{FranzTesanovic2000,VafekMelikyan2006}
$\mathcal{U}=e^{\frac{i}{2}\sigma_3\phi_{\br}}$ where the Pauli
sigma matrices act on the Nambu spinors. The transformed Hamiltonian
$\mathcal{H}(\bk)=e^{-i\bk\cdot\br}\mathcal{U}^{-1}\;\mathcal{\hat{H}}_0\;\mathcal{U}e^{i\bk\cdot\br}$
becomes
\begin{equation}\label{H0singgauge}
\mathcal{H}(\bk)=\sigma_3\left(
\tilde{\mathcal{E}}_{\br}(\bk)-\mu\right)+\sigma_1\tilde{\Delta}_{\br}(\bk),
\end{equation}
where the transformed lattice operators satisfy
\begin{eqnarray}
\mathcal{\tilde{E}}_{\br}(\bk)\psi_{\br}&=&-t\!\!\!\sum_{\bdelta=\pm\hat{\bx},\pm\hat{\by}}
z_{2,\br\br+\delta}\times e^{i\sigma_3 V_{\br\br+\delta}} e^{i\bk\cdot\delta}\psi_{\br+\delta}\\
\tilde{\Delta}_{\br}(\bk)\psi_{\br}&=&\Delta_{0}\!\!\!\sum_{\bdelta=\pm\hat{\bx},\pm\hat{\by}}
z_{2,\br\br+\delta}\times
\eta_{\bdelta}e^{i\bk\cdot\delta}\psi_{\br+\delta}.
\end{eqnarray}
The physical superfluid velocity enters via the factor
\begin{eqnarray}
 e^{i
V_{\br\br'}}
=\frac{1+e^{i(\phi_{\br'}-\phi_{\br})}}{|1+e^{i(\phi_{\br'}-\phi_{\br})}|}e^{-i\bA_{\br\br'}}
\end{eqnarray}
and represents the lattice analog of the semiclassical (Doppler)
effect. The Z$_2$ field $z_{2,\br\br'}=1$ on each bond except the
ones crossing the branch cut where $z_{2,\br\br'}=-1$.

If the vortices form a regular lattice, the transformed Hamiltonian
(\ref{H0singgauge}) is invariant under discrete translations by the
primitive vectors $\bR_1$ and $\bR_2$ defining the magnetic unit
cell, reflecting the periodicity of $V_{\br\br'}$ and the periodic
choice of the branch cuts. Consequently, it can be diagonalized in
the Bloch basis. By the Bloch condition $\mathcal{H}(\bk)$ acts on
the periodic functions, and the crystal wavevector $\bk$ varies
continuously within the 1$^{st}$ Brillouin zone defined by the
primitive reciprocal lattice vectors $\bK_1=2\pi
\frac{\bR_2\times\hat{\bz}}{\hat{\bz}\cdot(\bR_1\times \bR_2)}$,
$\bK_2=2\pi \frac{\hat{\bz}\times\bR_1}{\hat{\bz}\cdot(\bR_1\times
\bR_2)}$.

Note that, since the unitary transformation $\mathcal{U}$ is time
independent, the Hamiltonians (\ref{bdg0}) and (\ref{H0singgauge})
have the same thermodynamic and tunneling density of states.

In presence of the particle-hole symmetry ($\mu=0$), the Hamiltonian
(\ref{H0singgauge}), which is a $2n-$dimensional Hermitian matrix,
connects only sites belonging to different sublattices and can be
transformed into the form
\begin{eqnarray}\label{hWenZee}
\mathcal{H}(\bk)=\left(\begin{array}{cc} 0 & \mathcal{T}(\bk) \\
\mathcal{T}^{\dagger}(\bk) & 0 \end{array}\right).
\end{eqnarray}
Following Wen and Zee\cite{WenZee1989,WenZee2002} (see also
Ref.\cite{Volovik2006unpublished}), at all points $\bk$ where the
determinant of $\mathcal{H}(\bk)$ does not vanish, we can define
\begin{eqnarray}
\mathcal{M}(\bk)=\frac{\mathcal{H}(\bk)}{|\det
\mathcal{H}(\bk)|^{\frac{1}{2n}}}=
\left(\begin{array}{cc} 0 & h(\bk) \\
h^{\dagger}(\bk) & 0 \end{array}\right),
\end{eqnarray}
where $|\det h(\bk)|=1$. Therefore, if the phase of $\det h(\bk)$
winds by a nonzero integer multiple of $2\pi$ around a loop in
$\bk$-space, then $\det h(\bk^*)=0$ for some $\bk^*$ inside the loop
and $\mathcal{H}(\bk^*)$ has zero eigenvalues. The associated
winding number $m$, can be calculated from
\begin{eqnarray}\label{windingNumber}
m=\frac{1}{2\pi}\oint d\bk \cdot \ba
\end{eqnarray}
where the $\bk$-space vector field
\begin{eqnarray}\label{adef}
a_{\mu}(\bk)=-i\partial_{k_{\mu}}\ln
\det(h(\bk))=-\frac{i}{2}Tr\left[\gamma_5\mathcal{M}^{-1}\partial_{k_{\mu}}\mathcal{M}\right].
\end{eqnarray}
Since $\sigma_2 \mathcal{H}^{\ast}(\bk)
\sigma_2=-\mathcal{H}(-\bk)$, we find that the vector field is even
under $\bk$-space inversion, $a_{\mu}(\bk)=a_{\mu}(-\bk)$. Therefore
the circulation of $\ba$ vanishes at four $\bk$ points:
$\bk^*=\{0,\frac{1}{2}\bK_1,\frac{1}{2}\bK_2,
\frac{1}{2}(\bK_1+\bK_2)\}$. We thus arrive at a rather general
result, namely {\it if there are zero energy eigenstates at any of
these four points, they must appear in degenerate pairs with
opposite $m$, thereby canceling the circulation of $\ba$.} Such
situation is generally unstable to either nucleation of $m=+1$ and
$m=-1$ pairs with two massless Dirac fermions, or a vacuum of
$\bk$-space vortices implying a spectral gap.

To proceed with a specific example, consider vortex lattice with two
vortices per magnetic unit cell respecting site centered inversion
symmetry. The space inversion symmetry is realized as a projective
symmetry and the invariant gauge group\cite{WenZee2002} is $Z_2$,
i.e. ordinary space inversions $\mathcal{I}$ must be followed by
$Z_2$ gauge transformations $\gamma_{\br}$ \cite{VafekMelikyan2006}.
By constructing an operator
$\mathcal{P}=(-1)^{x+y}\gamma_{\br}\mathcal{I}$ which anticommutes
with $\mathcal{H}(\bk)$ at the four $\bk^*$ points, one can show
that the spectrum is gapless at these points with $2$ degenerate
Dirac cones at each $\bk^*$ giving the total of $8$ massless Dirac
fermions in the 1$^{st}$ Brillouin zone \cite{VafekMelikyan2006}.

What happens if the site centered inversion symmetry is broken?
According to the above general classification, either the spectrum
is gapped, or the $\bk$-space vortex-antivortex pairs are nucleated
resulting in two massless Dirac fermions. As shown below, the latter
case is realized for weak breaking of the site centered inversion
symmetry.

Without loss of generality, let us concentrate on $\mathcal{H}(0)$.
Since $\mathcal{H}(0)\mathcal{P}=-\mathcal{P}\mathcal{H}(0)$ and
since $Tr\mathcal{P}=4$, the four zero energy eigenstates of
$\mathcal{H}(0)$ are also eigenstates of $\mathcal{P}$ with the same
eigenvalue\cite{VafekMelikyan2006} of $+1$. It is also
straightforward to see that two of them can be chosen to be even
under the projective inversion $\gamma_{\br}\mathcal{I}$ and the
remaining two to be odd. The nullspace basis is then
\begin{eqnarray}\label{eq:nullspace}
\mathcal{A}:\left\{|1\rangle,|2\rangle=i\sigma_2|1\rangle^* \right\}\ldots even \\
\mathcal{B}:\left\{|3\rangle,|4\rangle=i\sigma_2|3\rangle^*
\right\}\ldots odd,
\end{eqnarray}
where the first set is even under $\gamma_{\br}\mathcal{I}$ and the
second one is odd. Since these states are $+1$ eigenstates of
$\mathcal{P}$, they are also eigenstates of $(-1)^{x+y}$, with the
eigenvalue $+1$ for the even ones and $-1$ for the odd ones. This
means that states which belong to the representation $\mathcal{A}$
are non-zero only on one sublattice, while the states which belong
to the representation $\mathcal{B}$ are non-zero on the
complementary one.

The dispersion of the energy near the degenerate quadruplet can be
found within degenerate perturbation theory. To the first order, the
low energy effective Hamiltonian is given by the nullspace matrix
elements
\begin{eqnarray}\label{effHam}
\mathcal{H}^{eff}_{mn}(\bk;\bu_1,\bu_2)=\left\langle
m\left|\mathcal{H}(\bk;\bu_1,\bu_2) \right|n\right\rangle,
\end{eqnarray}
where we explicitly included the dependence of the Hamiltonian on
$\bu_j$'s which denote the possible small displacements of the
positions of the two vortices within the magnetic unit cell. Due to
its particle hole symmetry, $\mathcal{H}(\bk;\bu_1,\bu_2)$ cannot
mix states with the same parity and therefore
$\mathcal{H}^{eff}_{mn}$ is block off-diagonal with the structure
(\ref{hWenZee}). By using $\sigma_2 \mathcal{H}^{\ast}(\bk)
\sigma_2=-\mathcal{H}(-\bk)$, one can show that
$\mathcal{H}^{*eff}_{13}(\bk;\bu_1,\bu_2)=-\mathcal{H}^{eff}_{24}(-\bk;\bu_1,\bu_2)$
and
$\mathcal{H}^{*eff}_{14}(\bk;\bu_1,\bu_2)=\mathcal{H}^{eff}_{23}(-\bk;\bu_1,\bu_2)$.
In addition, the projective inversion $\gamma_{\br}\mathcal{I}$
guarantees that
\begin{eqnarray}\label{dodd}
\mathcal{H}^{eff}_{mn}(\bk;\bu_1,\bu_2)=-\mathcal{H}^{eff}_{mn}(-\bk;-\bu_2,-\bu_1).
\end{eqnarray}

Clearly, for $\bu_j=0$ we have two degenerate gapless Dirac fermions
at $\bk=0$ and vanishing circulation of $\ba$ (\ref{windingNumber})
as required by the above argument.

For finite $\bu_j$ the zero energy states exist provided that
\begin{eqnarray}\label{det0firstorder}
\det\mathcal{H}^{eff}(\bk;\bu_1,\bu_2)=0
\end{eqnarray}
Due to (\ref{dodd}), we can expand near $\bk=0$, $\bu_j=0$ to find
$\mathcal{H}^{eff}_{13}(\bk;\bu_1,\bu_2)=k_{\mu}A_{\mu}+u_{j\mu}a^{j}_{\mu}$
and
$\mathcal{H}^{eff}_{14}(\bk;\bu_1,\bu_2)=k_{\mu}B_{\mu}+u_{j\mu}b^{j}_{\mu}$.
Note that $a_{\mu}^{1}=a_{\mu}^{2}=a_{\mu}$ and
$b_{\mu}^{1}=b_{\mu}^{2}=b_{\mu}$. There are two solutions to the
Eq.(\ref{det0firstorder}), at
$$k^{(c)}_{\mu}=\pm\sqrt{\frac{|u_{i\alpha} a^{i}_{\alpha}|^2+|u_{i\alpha}
b^{i}_{\alpha}|^2}{|A_{\alpha}\Lambda^{i}_{\alpha\beta}u_{i\beta}|^2+|B_{\alpha}\Lambda^{i}_{\alpha\beta}u_{i\beta}|^2}}
\Lambda^j_{\mu\nu}u_{j\nu},$$ where
$\Lambda^j_{\mu\nu}=2\varepsilon_{\mu\lambda}\Im
m\left[A_{\lambda}{a^j}^{*}_{\nu}+B_{\lambda}{b^j}^{*}_{\nu}\right]$
is independent of $\bu_j$. We see that small vortex displacements
induce a $\bk$-space vortex-antivortex pair, resulting in a pair of
gapless fermions near each $\bk^*$. By going to the second order
degenerate perturbation theory, I have checked that the character of
the solution does not change. {\it This is a result of the general
topological structure of $\mathcal{H}(\bk)$: the spectrum is gapless
and the number of the Dirac fermions is invariant under a finite
range of periodic perturbations which can break site centered
inversion, but do not break particle-hole symmetry.} In other words,
any p-h symmetric perturbation, such as supercurrent variations and
bond or pair density waves, must be sufficiently strong to cause the
annihilation of the $\bk$-space vortex-antivortex pair.

We now turn to the analysis of the magnetic field dependence of the
effective Hamiltonian (\ref{effHam}). According to the scaling
argument of Simon and Lee\cite{SimonLee1997}, since the vortex
displacement $\bu_j$ has a dimension of length, in the limit of
large $\ell$ we should have\cite{Marinelli2000}
$$
\mathcal{H}^{eff}_{mn}(\bk;\bu_j;\ell)=\frac{1}{\ell}\mathcal{F}^{(SL)}_{mn}\left(\ell\bk;\frac{\bu_j}{\ell}\right)
$$
where $\mathcal{F}^{(SL)}$ is the scaling function. As a result the
effective velocity of the gapless Dirac fermions should be $\ell$
independent, and in addition, the vortex displacement fields $\bu_j$
should enter the effective Hamiltonian as $\bu_j/\ell^2$, making
$a^j_{\mu}\sim b^j_{\mu}\sim \ell^{-2}$.

Remarkably, explicit numerical evaluation of the matrix elements
(\ref{effHam}) shows that while the effective velocity is indeed
$\ell$-independent\cite{VafekMelikyan2006}, the vortex displacements
coupling to the gapless fermions is much stronger and instead
$a^j_{\mu}\sim b^j_{\mu}\sim \ell^{-1}$. The correct scaling form is
rather
\begin{eqnarray}
\mathcal{H}^{eff}_{mn}(\bk;\bu_j;\ell)=\frac{1}{\ell}\mathcal{F}^{(anom)}_{mn}\left(\ell\bk;\frac{\bu_j}{\xi_0}\right)
\end{eqnarray}
where $\xi_0$ is a short distance cutoff of the order of the vortex
core radius. This is a consequence of a quantum
anomaly\cite{Jackiw1972,Camblong2001} which arises from two effects:
first, due to the $1/r$ increase of the superfluid velocity near a
vortex, the extended quasiparticle wavefunctions experience an
anomalous
increase\cite{VafekMelikyan2_2001,Melikyan2006,Melikyan2006unpublished}
$\sim1/\sqrt{r\ell}$ cut off only by a short distance scale $\xi_0$.
Second, the potential due to the displaced vortex, as experienced by
the nodal quasiparticle of the unperturbed vortex lattice, is
equivalent to the potential due to vortex-antivortex "dipole". At
distances much larger than $|\bu_j|$, this "dipole" potential falls
off as $\sim1/r^2$. The resulting matrix element scales as
$|\bu_j|/(\ell\xi_0)$, rather than $|\bu_j|/\ell^2$. The scale
invariance of the linearized Dirac equation\cite{SimonLee1997} is
thus broken and just as in the theory of critical phenomena, the
short distance cutoff effects long distance properties via an
anomalous dimension. In the case at hand, the anomalous dimension of
$\bu_j$ is effectively $1$.

In the weakly disordered vortex lattice, the disorder-free vortex
lattice nodal quasiparticles experience a random potential produced
by the vortex dipoles with random orientation and strength. The
vortex displacements $\bu_{\bR_j}$, corresponding to the vortex
position $\bR_j$, can be assumed to be uncorrelated
\begin{eqnarray}
\langle
u_{\alpha\bR_i}u_{\beta\bR_j}\rangle=\bu^2_{rms}\delta_{ij}\delta_{\alpha\beta}.
\end{eqnarray}
The effective Hamiltonian near $\bk=0$ is $
\mathcal{H}^{eff}_{\bk,\bk'}=$
\begin{eqnarray}\delta_{\bk\bk'}k_{\mu}\left(\begin{array}{cc}0
& v_{\mu}\\
v^{\dagger}_{\mu} & 0\end{array}\right)
+\frac{\ell^2}{L^2}\sum_{j}u_{\bR_j\mu}e^{-i(\bk-\bk')\cdot\bR_j}\left(\begin{array}{cc}0
& \lambda_{\mu}\\
\lambda^{\dagger}_{\mu} & 0\end{array}\right)
\end{eqnarray}
where $v_{\mu}=\left(\begin{array}{cc}A_{\mu} & B_{\mu}\\
-B^*_{\mu} & A^*_{\mu}\end{array}\right)$,
$\lambda_{\mu}=\left(\begin{array}{cc}
a_{\mu} & b_{\mu}\\
b^*_{\mu} & -a^*_{\mu}
\end{array}\right)$,
and the sum is over the vortex positions. $L^2$ is the area of the
system, and just as before, $a_{\mu}\sim b_{\mu}\sim \ell^{-1}$.

Near $\bk=0$, we can treat the vortex-antivortex dipole scattering
within the self-consistent Born approximation and the quasiparticle
(retarded) self-energy
\begin{eqnarray}
\Sigma(\omega)=2\bu^2_{rms}\left[|a|^2+|b|^2\right]\ell^2\!\!\int\!\!\frac{d^2\bq}{(2\pi)^2}
\mathcal{G}_{\bq}^{(0)}(\omega-\Sigma(\omega))
\end{eqnarray}
where the integral is over the magnetic Brillouin zone, and
$|a|^2+|b|^2=\sum_{\mu}(|a_{\mu}|^2+|b_{\mu}|^2)$;
$\mathcal{G}_{\bq}^{(0)}$ is the quasiparticle Greens function for
the disorder free vortex lattice.

We find the vortex disorder scattering rate, $\tau_{v}^{-1}$, from
the imaginary part of the retarded self-energy to be
\begin{eqnarray}\label{scatRate}
\frac{1}{\tau_v}=-\Im
m\Sigma(0)\approx\frac{\pi\sqrt{v_{\parallel}v_{\perp}}}{\ell}\exp\left\{
\frac{-\pi
v_{\parallel}v_{\perp}}{\ell^2(|a|^2+|b|^2){\bu^2_{rms}}}\right\}
\end{eqnarray}
where $v_{\parallel,\perp}$ are the velocities of the dispersion of
the (disorder-free) vortex lattice Dirac quasiparticles along the
principal axes. Note, that due to the quantum anomaly, the
scattering rate has the scaling form
$$\tau_{v}^{-1}=\sqrt{H}\;\Gamma\!\left(\frac{\bu^2_{rms}}{\xi_0^2}\right).$$
This should be contrasted with the result obtained from the naive
scaling argument
$\tau_{(SL)}^{-1}=\sqrt{H}\Gamma\left(\bu^2_{rms}H/\Phi_0\right)$,
which is parametrically smaller than (\ref{scatRate}).

The contribution to the zero energy density of states from the
vicinity of $\bk=0$ is therefore
$$N_{\bk=0}(0)=\frac{2}{\pi^2v_{\perp}v_{\parallel}\tau_v}\ln\left(\pi\tau_v\sqrt{v_{\parallel}v_{\perp}}\ell^{-1}\right)$$

Upon summing the contribution from all four nodes, we find that that
the vortex disorder induced density of states has a scaling form
$N(0)=\sqrt{H}\rho(\bu_{rms}^2/\xi^2)$.

Since the vortex disorder induced self-energy was argued to be given
by the self-consistent Born approximation, in order to find the
electronic contribution to the thermal conductivity $\kappa_{xx}$ we
need to calculate the vertex correction by summing the ladder
diagrams. In the limit of small $\bu_{rms}$ we find
\begin{eqnarray}
\frac{\kappa_{xx}}{T}=\frac{\pi^2}{3}\frac{k_B^2}{\hbar}\frac{2}{\pi^2}\left[\frac{v_{\parallel}}{v_{\perp}}+\frac{v_{\perp}}{v_{\parallel}}\right]
\beta_{V},
\end{eqnarray}
where $\beta_{V}=1+\mathcal{O}(\bu_{rms}^4)\gtrsim 1$.

The above result is independent of magnetic field and depends only
on the vortex lattice renormalized Dirac cone anisotropy. The extra
factor of $2$ relative to Ref.\cite{DurstLee2000} comes from the
node doubling \cite{VafekMelikyan2006}.

In conclusion, I showed that weak vortex lattice disorder exhibits
an anomalously strong coupling to the low energy fermions, which
violates Simon-Lee scaling in a novel way. The resulting vortex
disorder induced density of states was calculated for weak disorder
and found to scale as $\sqrt{H}f(\bu^2_{rms}/\xi^2)$. The electronic
contribution to $\kappa_{xx}/T$ was found to be magnetic field
independent, but different from its $H=0$ value. The latter can be
understood to result from the cancelation of the $H$-dependence of
the vortex disorder induced density of states and the mean free
path, which is set by the intervortex separation in this ultraclean
limit. This result is in qualitative agreement with the experimental
finding on the ultraclean YBCO\cite{Hill2004}, although quantitative
agreement is unlikely due to the significant vortex disorder in the
actual sample. Nevertheless, the theory presented here, while
strictly limited to the case of weak positional vortex disorder,
does suggest that as $\bu_{rms}$ increases and becomes of order of
the magnetic length $\ell$, the vortex disorder induced density of
states continues to scale as $\sqrt{H}$. Since the mean free path is
still set by $\ell$, this suggests $H$-independent, albeit vortex
disorder dependent, $\kappa_{xx}$.

I wish to thank Dr. A. Melikyan for numerous useful discussions as
well as Prof. Z. Tesanovic for critical reading of the manuscript.
\bibliography{d}
\end{document}